\documentstyle[12pt,fleqn]{article}
\def\be{\begin{equation}}
\def\ee{\end{equation}}
\def\ba{\begin{eqnarray}}
\def\ea{\end{eqnarray}}

\def\fun#1#2{\lower3.6pt\vbox{\baselineskip0pt\lineskip.9pt

\ialign{$\mathsurround=0pt#1\hfill##\hfil$\crcr#2\crcr\sim\crcr}}}
\def\ref#1{$^{#1)}$}
\begin{document}
\begin{titlepage}
\begin{center}
\today    \hfill       LBL- 34827\\
            \hfill       UCB-PTH-93/31\\
{\bf Hypercharge and the Cosmological Baryon Asymmetry. }\footnote{This
work was supported in part by the Director, Office of
Energy Research, Office of High Energy and Nuclear Physics, Division of
High Energy Physics of the U.S. Department of Energy under Contract
DE-AC03-76SF00098 and in part by the National Science Foundation under
    grant PHY-90-21139.}
\vskip .25in
      {\bf Aram Antaramian \\
           Lawrence J. Hall \\
           Andrija Ra\v{s}in}\footnote{On leave of absence from the Ruder
           Bo\v{s}kovi\'{c} Institute, Zagreb, Croatia.}\\
 \medskip
{\em  Theoretical Physics Group\\
      Lawrence Berkeley Laboratory\\
      University of California\\
      Berkeley, CA 94720}
       \end{center}

\begin{abstract}
Stringent bounds on baryon and lepton number violating interactions have
been derived from the requirement that such interactions, together
with electroweak instantons, do not destroy a cosmological baryon asymmetry
produced at an extremely high temperature in the big bang. While these bounds
apply in specific models, we find that they are generically evaded. In
particular, the only requirement for a theory to avoid these bounds is that it
contain charged particles which, during a
certain cosmological epoch, carry a non-zero hypercharge asymmetry. Hypercharge
neutrality of the universe then dictates that the remaining
 particles must carry a
compensating hypercharge density, which is necessarily shared amongst them
so as to give a baryon asymmetry. Hence the generation of a
hypercharge density in a sector of the theory forces the universe to have a
baryon asymmetry.

PACS number(s): 05.20.-y, 11.30.-j, 98.80.Cq
\end{abstract}
\end{titlepage}

\newpage

\begin{center}
{\bf Disclaimer}
\end{center}

\vskip .2in

\begin{scriptsize}
\begin{quotation}
This document was prepared as an account of work sponsored by the United
States Government.  Neither the United States Government nor any agency
thereof, nor The Regents of the University of California, nor any of their
employees, makes any warranty, express or implied, or assumes any legal
liability or responsibility for the accuracy, completeness, or usefulness
of any information, apparatus, product, or process disclosed, or represents
that its use would not infringe privately owned rights.  Reference herein
to any specific commercial products process, or service by its trade name,
trademark, manufacturer, or otherwise, does not necessarily constitute or
imply its endorsement, recommendation, or favoring by the United States
Government or any agency thereof, or The Regents of the University of
California.  The views and opinions of authors expressed herein do not
necessarily state or reflect those of the United States Government or any
agency thereof of The Regents of the University of California and shall
not be used for advertising or product endorsement purposes.
\end{quotation}
\end{scriptsize}

\vskip 2in

\begin{center}
\begin{small}
{\it Lawrence Berkeley Laboratory is an equal opportunity employer.}
\end{small}
\end{center}

\newpage
\setcounter{page}{1}

\section {Introduction}

Various authors \cite{Fukugita,Nelson} have placed cosmological
bounds on the size of baryon and lepton number violating interactions
in theories where baryogenesis occurs before the electroweak phase transition.
The baryon asymmetry of the universe is threatened by a
combination of these interactions and a
 large electroweak instanton rate \cite{Hooft,Kuzmin}.
Electroweak instanton interactions are expected to be in
equilibrium for temperatures above $T_{min}$, approximately the weak
breaking scale, up to some very high temperature $T_{max}\simeq 10^{12}GeV$.
Such reactions create $SU(2)_L$ transforming fermions out
of the vacuum \cite{Hooft}.
 Lepton and baryon violating
interactions such as R-parity breaking terms in supersymmetry \cite{Hall1} or
Majorana neutrino masses, when in
equilibrium simultaneously with instanton reactions can break all linear
combinations of conserved quantum numbers which involve
baryon  number. Naively, one is led to believe that the baryon
asymmetry of the universe is therefore washed away. In this paper
we examine the general circumstances in which this
outcome is avoided.
We find that in many
models there will be additional symmetries and, even though these symmetries
apparently have nothing to do with baryon number, they
automatically lead to a protection of it.

It is well known that a symmetry which involves baryon number itself,
such as  $B-3L_{i}$, can preserve the baryon asymmetry \cite{Nelson}.
Approximate symmetries involving $B$ have been found in the minimal
supersymmetric standard model which can be used to help prevent erasure of the
baryon asymmetry \cite{Ibanez}. We have found that the
protection of the baryon asymmetry is extremely common and is a typical feature
of theories with extra symmetries, even when those symmetries do not transform
quarks. We illustrate this by a very simple example: assume that there exists
a particle, $X$, which carries hypercharge but not SU(2) or SU(3) gauge
interactions\footnote{Cline, Kainulainen, and Olive have recently
employed the right handed electron as
the $X$ particle \cite{cline}. However, in this case non-zero
Yukawa interactions
limit the temperature range  over which the right handed electron
can maintain a hypercharge asymmetry.}.
 Assume that reactions occurring at temperatures well above
$T_{min}$
generate an asymmetry in the X species, and that at lower temperatures
the reactions which change $X$ number are sufficiently
weak that this $X$ asymmetry persists.  A crucial
role is played by the requirement that the early
universe is hypercharge neutral.
Because $X$ particles carry hypercharge, the  asymmetry in their number
contributes to the hypercharge density of the universe. The remaining
particles in the theory must carry an opposite hypercharge density to cancel
this. Chemical equilibrium equations
specify how this hypercharge density is shared. A baryon asymmetry can develop
either through added $B$ violating interactions or once the
weak instanton becomes effective.  In general
any $X$ asymmetry together with chemical equilibrium requires
a baryon asymmetry\footnote{Implicit in this
discussion is the assumption that the universe is homogeneous.}.
This illustrates just how easy it is to preserve the
baryon asymmetry and, to our way of
thinking, puts the issue of direct detection of baryon and lepton number
violation back where it belongs: with the experimentalists.

\section{General condition for survival of a baryon asymmetry.}

In this section we discuss, in a very general way, the conditions under which
an extra $U(1)$ symmetry preserves the cosmological baryon asymmetry.

      In thermodynamic equilibrium the number density of particle
species $i$ is determined by its chemical potential, $\mu_{i}$. If a given
reaction, say $p_{1} + p_{2} \rightleftharpoons p_{3} + p_{4}$, is in
equilibrium then $\mu_{1} + \mu_{2} = \mu_{3} + \mu_{4}$.
It is straightforward, yet tedious, to solve all chemical equilibrium
equations. One can simplify the process by noticing that these equations
are the same equations one would write down to determine
the $U(1)$ symmetries of the equilibrium theory. One need only replace
$\mu_{i}$ with $q_{i}$, the charge of particle $i$. Solving
for $q_{i}$ determines the possible assignments of $U(1)$
charge to each particle so
that all equilibrium reactions conserve that
charge.  In general such $U(1)$ symmetries need not be exact symmetries of the
Lagrangian. They are symmetries of those interactions in thermal equilibrium
at temperature T, and we refer to them as effective $U(1)$ symmetries at this
temperature.

Thus, a solution to
the chemical equilibrium equations corresponds to an assignment of
effective $U(1)$ charges to each particle, and the possible effective $U(1)s$
in a given theory are usually easy to identify. Suppose that at a certain
temperature there are N such effective U(1)s: $U(1)_A , A = 1,... N$, then
the most general solution is
$$
\mu_{i} = \sum_{A} C^A q^A_i      \eqno(1)
$$
where $q^A_i$ is the charge of particle $i$ under $U(1)_A$. The constant
$C^A$ we refer to as the asymmetry constant for $U(1)_A$. As soon as some
interaction which violates $U(1)_A$ comes into thermal equilibrium, $C^A$
rapidly tends to zero:  $U(1)_A$ is no longer able to support particle
asymmetries.

This general solution is restricted, however. We assume that the universe is
homogeneous and that no charge asymmetry has
developed for the unbroken gauged $U(1)s$ of the theory\footnote{This
requirement avoids the problems inherent in giving a massless gauge
boson a chemical potential.}. This forces the
charge density for these $U(1)s$ to zero. We can write the charge
density for $U(1)_A$ as follows,
$$
Q^A = \sum_{i} q^A_i \: n_i   \eqno(2)
$$
where $n_{i}$ is the particle asymmetry density of species $i$.
If particle asymmetry densities are small then they
can be written, for $T\gg m_i$, as
$$
n_{i} \simeq  \frac{T^2} {6} \tilde{g}_i \mu_i    \eqno(3)
$$
where $\tilde{g}_i$ is the number of internal degrees of freedom of particle
$i$,  $g_i$, multiplied by a factor of two for bosons.
(However, see reference \cite{Ross} for an interesting look at small mass
effects.)
Under these conditions the charge density constraint is a simple linear
equation in the $\mu_{i}$s.
$Q^A$ can be written using $n_i$ from (3) and $\mu_i$ from (1):

$$
Q^A \simeq {T^2\over 6} \sum_B C^B \sum_i \widetilde{g}_i q^B_i q^A_i
    = {T^2\over 6} \sum_B C^B \; \overline{B}\cdot\overline{A}\eqno(4)
$$

where we define $\overline{B}\cdot\overline{A}$ by
$$
\overline{B}\cdot\overline{A} = \sum_i \widetilde{g}_i q^B_iq^A_i.\eqno(5)
$$

Should the diagonal generators of non-Abelian gauge groups, such as $T_{3L}$, \
be included in the list of effective $U(1)s$?
The answer is no, as can be seen easily from the above equations.
Call such a generator $\alpha$, then neutrality of the universe with respect to
this charge requires
$$
\sum_A C^A\overline{A}\cdot\overline{\alpha} =0\eqno(6)
$$
where
$$
\overline{A} \cdot \overline{\alpha} = \sum_i \widetilde{g}_i q^A_i
q^\alpha_i.\eqno(7)
$$
When $A$ refers to a $U(1)$ generator (not embedded in a non-Abelian gauge
group) then $\overline{A}\cdot \overline{\alpha} =0$.
This is because the $\widetilde{g}_i$ and $q^A_i$ are the same for all
components of an irreducible representations of $\alpha$, and hence the sum in
(7) can be written as a sum of zero terms, one for each irreducible multiplet
of $\alpha$.
When $A=\beta$ is a diagonal generator of a non-Abelian group the
orthogonality property of the generators within each multiplet ensures that
$\sum_i q^\alpha_i q^\beta_i$ vanishes for $\beta \neq \alpha$.
Hence the sum  in (6) just has one term: $C^\alpha \overline{\alpha}\cdot
\overline{\alpha} =0$.
Since $\overline{\alpha}\cdot \overline{\alpha} \neq 0$, we have proved that
$C^\alpha =0$ follows from $Q^\alpha =0$.
This implies that such $U(1)s$ need not be included in the list of effective
$U(1)s$.

Now let's apply this formalism. We are interested in the situation in
which  additional particles and interactions have been  added to the standard
model such that at temperatures $T$, $T_c < T < T_{max}$, where $T_c$ is the
weak breaking temperature, there are just two
effective $U(1)s$: $Y$ and $X$, where  $Y = 2(Q-T_{3})$ denotes hypercharge
and $X$ is an ungauged effective symmetry.
The charge neutrality condition
 (6) when applied to  hypercharge gives
$$
C^Y =- {\overline{X}\cdot\overline{Y}\over \overline{Y}^2} C^X.\eqno(8)
$$
Using (8) in equation (4) the asymmetry in baryon number is just
$$
n_B \simeq {T^2\over 6} C^X \left(\overline{X} - {\overline{X} \cdot
\overline{Y}\over\overline{Y}^2} \overline{Y}\right) \cdot
\overline{B}.\eqno(9)
$$
where we have rewritten $Q^{B}$, the baryon density, as $n_B$.
This is the general result of this paper.
Any effective $U(1)_X$, whether it contains a piece of baryon number or not,
will in general contribute to $n_B$ if $C^X \neq 0$.
The extension of (9) to many extra $X$ symmetries is straightforward.
Providing such a $U(1)_X$ exists, there is no limit to how large the $B$ and
$L$ violating interactions can be.

We will examine the case in which $X$ particles carry no baryon number
themselves. Then $$
n_{B} \simeq {T^{2}\over6} C^{X}
\left( - {\overline{Y} \cdot \overline{B}\over \overline{Y}^{2}} \right)
  \overline{X}\cdot \overline{Y}. \eqno(10)
$$

In the standard model
$\overline{Y}\cdot \overline{B}/\overline{Y}^2 ={1\over11}$.
Additional particles will change this, but would generally give some non-zero
value which we call $\alpha$.
Then $n_B \simeq - {T^2 \over6}\alpha C^X  \left( \overline{X}\cdot
\overline{Y} \right)$.
Thus to obtain $n_B \neq 0$ we require   that some particles with
$X_i \neq 0$ have $Y_i
\neq 0$.
Hypercharge neutrality then forces other particles to have an asymmetry, some
of which carry baryon number, thus providing a baryon asymmetry.

Cline {\em et al.} \cite{cline} point out that in the standard model right
handed electron number is conserved down to a temperature of about
$10 \: TeV$, and thus can insure  that baryon/lepton violating interactions
don't wash away the baryon asymmetry above this temperature. Thus the
standard model already contains  $X$ particles in the form of right
handed electrons. In section 3, we discuss another possibility, an
$X$ symmetry which does not
transform any standard model particles. In
this case  (8) can be rewritten
in terms of the hypercharge density carried by the standard model sector,
$Q^{Y}(SM)$, and
by the $X$ sector, $Q^{Y}(X) \equiv \sum_{i} q_{i}^Y n_{X_i}$.

$$
Q^{Y}(SM) = Q^{Y}(X)\eqno(11)
$$

In terms of $Q^{Y}(X)$ equation (10) becomes
$$
n_{B} \simeq - {1\over11} Q^{Y}(X). \eqno(12)
$$
(We have assumed that $T< 10 TeV$ so that right handed electrons
are in equilibrium.)

Equation (12) doesn't assume that $X$ number density is small or
proportional to its chemical
potential. Thus it is valid even when the temperature drops below the
mass of certain $X$ particles.
When this happens the heavier species carrying $X$ might decay
into lighter ones.
Nevertheless, providing the particles with $X\neq 0$ possess a
hypercharge asymmetry the baryon asymmetry will survive.
In particular the $X\neq 0$ particles must continue to carry such an
asymmetry until a temperature $T_{0}$,
beneath which $B$ and $L$ violating reactions are sufficiently weak that a
symmetry having a baryon number component has become an effective
$U(1)$.
The  resulting baryon asymmetry after $X$ decay depends on the
specifics of the
model. In the least complicated scenario, in which baryon number is a good
symmetry below $T_0$, today's baryon asymmetry is simply derived form (12) and
entropy considerations.

We note that it is not necessary for our $X$
sector to be neutral under $SU(2)$. Adding additional
$SU(2)$ transforming fermions to the
standard model will mean that these particles also
take part in instanton mediated reactions.
Nevertheless, in a consistent theory, instanton reactions
will conserve the hypercharge asymmetry carried by the $X$ sector of the
theory. This is true because instantons neither violate hypercharge in
the standard model sector nor in the theory overall, and thus must
conserve hypercharge in the $X$-sector.

In this section we have tacitly assumed
 that some component of baryon number is a
good symmetry below $T_c$, the weak breaking temperature.
If this is not the case, then,
for temperatures $T$, $T_0<T<T_c$, the role of hypercharge is played by
electric charge. In this case the $X$ sector
must carry an electric charge asymmetry.

An intriguing possibility exists if the
lightest $X$ particle is stable and electrically neutral. If this is the case,
the particle is a
candidate for the dark matter in the universe \cite{Barr,Dod}.
To realize such a scenario, the $X$ sector would still have to
maintain a hypercharge asymmetry for temperatures above $T_0$. (For
convenience,
we have assumed $T_0 \geq T_c$.). However, at a
lower
temperature, charged $X$ particles would decay to standard model particles
plus these electrically neutral
$X$ particles.
If $ \Omega_{X}$ is the fraction of the critical density contributed
by the electrically neutral $X$ particles then their mass is given by
$$
{m_{X}\over m_{proton}} \simeq { \langle q_{X} \rangle \over11}
10^{2} \Omega_{X}
$$
where $\langle q_{X} \rangle$ is the appropriate   average of $X$-particle
hypercharges.
Low-back\-ground Ge detector experiments \cite{Sad,Ahlen} indicate that
an electrically neutral
 dark matter particle with nonzero hypercharge must have a mass
greater than $\sim 1000\: GeV$. Thus, we can effectively rule out a dark matter
$X$ particle with nonzero hypercharge.
One possible candidate is the  neutral component of a
new hyperchargeless $SU(2)$ multiplet. Such a particle is expected to interact
via loop diagrams with nuclei and thus its cross section
with $Ge$ is approximately
$10^{-35} cm^{2}$ or smaller \cite{Dod}, effectively evading
relevant experimental limits \cite{Sad}. Another candidate is a new
particle with no gauge interactions
whatsoever \cite{Barr}.

\section{A Simple Model.}

In this section we illustrate the general ideas discussed above with a very
simple model. We add to the standard model a single fermion X, of mass $m_X$,
which is SU(2) neutral but has three units of electric charge. It is unstable,
decaying to three charged leptons via the effective interaction
$$
\frac {1}{2}  \frac{1}{M^{2}} \;  \sum_{ijk} \;
    f_{ijk}\left( \overline{X} \: e^{i}_{R} \right)
  \; \left( (e^{j}_{R})^{T} Ce^{k}_{R}   \right) \:\:\: + \:\:\: h.c.
$$
where $e^{i}_{R}$ is the right handed lepton field of
flavor $i$, $X$ is the $X$ particle field,
$C$ is the charge conjugation matrix, $M$ is a
constant with units of energy, and $f_{ijk}$ ($=f_{ikj}$) is a flavor
dependent constant of order $1$. In addition we let our
model include unspecified
lepton and/or baryon violating terms which together
with the instanton reaction break all linear combinations of
$B$ and $L$ numbers.

 Both the mass of the $X$ particle, $m_{X}$, and the constant
$M$ are constrained by the various requirements of our theory. First
we must insure that the $X$ asymmetry develops before
all baryon violating interactions fall out of equilibrium. Otherwise the $X$
asymmetry has no effect on baryon number. Let $T_{X}$ be the
temperature at which $X$ violating reactions drop out of equilibrium.
Without specifying the exact scenario, we assume that an $X$ asymmetry
develops at some temperature lower than $T_{X}$ but above the
temperature at which instantons freeze out
(See \cite{Kolb} and references therein for numerous
methods by which number asymmetries can develop).  In this way the instanton
reaction provides the baryon violation required for our mechanism to
work.
This is a convenient choice, but not a necessary one
if other baryon violation exists in the theory.

It is interesting to note that the only baryon
violation required in this model is instantons. If an $X$ asymmetry exists or
develops during the epoch in which instantons are
in equilibrium then it will necessarily generate a
proportional baryon asymmetry.

In our example $X$ particles will eventually decay into standard model
particles. Various constraints must be imposed on this decay. To make
things  simple we require  $X$
particles to survive past the temperature at which instantons
freeze out.  We assume that after this temperature baryon
number is a good symmetry. Thus,
the only possible effect on the produced baryon density comes from the change
in entropy of the universe upon $X$ decay.

The standard nucleosynthesis scenario places limits
on this decay \cite{Kolb}. If $X$ particles
decay after nucleosynthesis, they
must not dump more than a factor of $\sim15$ times
the entropy density present at the time of nucleosynthesis.
If they did than the observed
baryon to photon density would be incompatible with standard
nucleosynthesis. Also, if the mass of the $X$
particle is larger than a few $MeV$, which it must be to avoid strict
limits on the width of the  $Z$ boson, then energetic photons from $X$
decay can destroy too much deuterium. Further, depending on the era of
decay,
 photons from $X$ decay can destroy the uniformity of the cosmic
microwave background radiation or contribute too much to the diffuse
photon background. If $X$ particles decay before nucleosynthesis, their
mass and density prior to decay must be compatible with the known baryon to
photon ratio, $\eta$, during nucleosynthesis.

Let us examine our first constraint. The rate for $X$ violating 4-fermion
interactions is given by
$$
 \Gamma_{X}   \simeq {49 f^{2} \pi^{5}\over 12960 \zeta(3)}  \; \frac
{T^{5}} { M^{4} }
$$
where $f^{2}$ is an average of terms like
$f_{ijk}^{\ast} f_{lmn}$, and we have dropped terms of order ${m_X \over T}$.

The Hubble constant,
$H$, is  $17 \frac{T^{2}}{M_{p}}$. The 4-fermion interaction
drops out of equilibrium when its rate falls below the Hubble
expansion rate\footnote{For convenience, and because we are
interested in the order of
magnitude of our results, we assume $g_{\ast} \simeq 106$
independent of the temperature.}. Calling the temperature
at which this occurs $T_{X}$, we have

$$
M^{4} \simeq {49 f^{2} \pi^{5}\over 220320 \zeta(3)} \; M_{p}
T_{X}^{3} .\eqno(13)
$$

Although $X$ number changing interactions freeze out at $T_{X}$,
$X$ particles stay in thermodynamic equilibrium below this temperature
through
their gauge interactions. These gauge interactions freeze out at a much lower
temperature given by the standard cold relic freeze out criteria.

Now we examine the decay of the $X$ particles.
The decay rate for these particles is given by
$$
\Gamma \simeq {f^{2} \over 256 \pi^{3}} \frac{m_{X}^{5}}{M^{4}}
$$
where we have ignored terms of order the
temperature over $m_{X}$ since they will be seen to be
negligible. The $X$ particles decay when this rate is
approximately equal to the Hubble expansion rate. Calling the temperature at
which these rates become equal $T_{D}$, we have
$$
M^{4} \simeq {f^{2} \over 4352 \pi^{3}}
\frac{m_{X}^{5}}{T_{D}^{2}} M_{p}. \eqno(14)
$$
If significant entropy is generated by $X$ decay then $T_{D}$ is the
``reheat'' temperature after decay.

Equations (13) and (14) can be combined to give
$$
m_{X}^{5} \simeq 7.6\times10^3 \;T_{X}^{3} T_{D}^{2}
$$

In figure 1 we plot the  allowed parameter space by considering the
constraints discussed above. (We have assumed $T_D \leq T_{min} \simeq 10^2
GeV$
and required $T_X >10 T_{min}$.)

The diagonal dotted lines in this figure are lines of
constant $T_X$ and are labeled in $GeV$.
The allowed region is divided up into
three regimes. The first, corresponding to $T_D > 10^{-3} GeV$,
covers the case in which $X$ particles decay before the onset of
nucleosynthesis.
In this case $X$ density just before decay may be
quite large, leading to an early matter dominated era and a significant
increase in entropy density upon $X$ decay. This is because
for large $m_X$, $X$ particle gauge interactions freeze out when
there is still a large anti-$X$ particle density.
In this situation, the $X$ number asymmetry is a small fraction
of the symmetric  relic freeze out density. A large
symmetric relic density leads
to large entropy dumping when $X$ particles decay. Let
us call the factor by which entropy
is increased $R$. Since, in our model, today's observed baryon
asymmetry is proportional to the $X$ asymmetry divided
by $R$, a large $X$ asymmetry is required
when $R$ is large.
We have plotted
a dot-dashed line which corresponds to the onset of
significant entropy generation when $X$ particles decay. At this line entropy
is increased by $10\%$ upon $X$ decay. As we rise above
this line the amount of entropy generated when the $X$ particles decay
increases. At the  top boundary of our allowed region
the $X$ asymmetry required to generate todays observed baryon asymmetry
becomes infinite. Above this line there is
no way to generate enough baryon asymmetry.

In the second regime $10^{-4}GeV < T_D < 10^{-3}GeV$,
during which nucleosynthesis is taking place we impose the conservative
requirement that $X$ decay increases the universe's entropy by
less than $10\%$. This is shown as a  dip in
the top boundary of the allowed region.

The last regime, $T_D < 10^{-4}GeV$, in which $X$
particles decay after nucleosynthesis, is bounded on the left
by the requirement that decay products don't destroy too
much deuterium \cite{Lindley}. The curved line marked with an arrow
takes account of this limit (We have used Lindley's
rough calculation for heavy dark matter particles \cite{Lindley}).
This constraint is more severe than those arising
from cosmic microwave background and diffuse photon background
observations. The top limit of this region is determined
by entropy dumping considerations. Since, in this case $X$ particles are
still present during nucleosynthesis, we know that the required
$X$ asymmetry is equal to ${-q_X \over 11}$ times
the baryon asymmetry at the time of
nucleosynthesis. When $X$ particles decay they can increase
the entropy and thus decrease the value of $\eta$ today relative to its
value during nucleosynthesis. We allow at most a decrease by
a factor of $15$, and this gives us our top limit.
 Figure 1 illustrates how general our mechanism is. The $X$
particle's mass can range
over $12$ orders of magnitude, from $45 \: GeV$ to $10^{12} \: GeV$.

\section{Conclusion}
We have shown that in order to avoid the strict
cosmological limits placed on lepton and baryon number
violating interactions it is not necessary to resort to low temperature baryon
generation or to the addition of new symmetries which affect baryons.
Any symmetry which allows one sector of the theory to
acquire a net hypercharge density
will suffice. This includes a
symmetry under which standard model particles are neutral,
as our example shows.
The key observation is
that, although this new symmetry seems decoupled from the
rest of the theory, the gauged $U(1)$
symmetries can connect it. Thus an asymmetry in $X$ particles, because
they are charged, is enough to ensure a proportional asymmetry in all
charged particles independent of whether their particle number
is conserved or not.

  If a  scenario similar to the one proposed here was realized in the
early universe, than experimental searches for lepton and baryon
violating interactions may prove successful. Such a success would not
only directly  signal exciting new  $L$ and/or $B$ number violating
physics, but would also indirectly signal the existence of a baryon
number protection mechanism.

\section{Acknowledgement}
This work was supported in part by the Director, Office of
Energy Research, Office of High Energy and Nuclear Physics, Division of
High Energy Physics of the U.S. Department of Energy under Contract
DE-AC03-76SF00098 and in part by the National Science Foundation under
    grant PHY-90-21139

\newpage
{\bf FIGURE CAPTION}

\noindent {\bf Figure 1:} The allowed parameter space in our example is shown,
bounded by solid lines. We have assumed $T_D \leq T_{min} \simeq 10^2 GeV$
and required $T_X >10 T_{min}$ and $m_X> 45 \; GeV$.
The diagonal dotted lines are lines of constant $T_D$, and are labeled in
$GeV$. Our parameter $M$ is also constant on these dotted lines,
$M = 2.9 \times 10^4 \left( {T_X \over GeV} \right)^{3/4} \: GeV$.
On the dot-dashed line the entropy of the universe
is increased by $10\%$ when $X$ particles decay. In determining this line as
well as the top boundary line we have assumed that
$X$ particle gauge interactions freeze out according to the
standard cold relic freeze out criteria \cite{Kolb}. We have made conservative
assumptions in determining the relative increase in entropy upon $X$ decay;
allowing the cosmic scale factor to scale as $t^n$ where $n$ ranges from $1/2$
to $2/3$. We have used a value for $\eta$ at the time of nucleosynthesis equal
to $\left( {11 \over 4} \right) 3\times10^{-10}$   .


\newpage
\begin{thebibliography}{99}

\bibitem{Fukugita} M. Fukugita and T. Yanagida, Phys. Rev. D {\bf42},
1285 (1990);
B. Campbell, S. Davidson, J. Ellis and K.A. Olive, Phys. Lett. B
{\bf256}, 457 (1991); Astropart. Phys. {\bf1}, 77 (1992);
J. A. Harvey and M. S. Turner, Phys. Rev. D {\bf42}, 3344 (1990).

\bibitem{Nelson} A.E. Nelson and S.M. Barr, Phys. Lett. B {\bf246}, 141 (1991);
W. Fischler, G.F. Giudice, R.G.
Leigh and S. Paban, Phys. Lett. B {\bf258}, 45 (1991).

\bibitem{Hooft} G. 't Hooft, Phys. Rev. Lett. {\bf37}, 8 (1976); Phys.
Rev. D {\bf14}, 3432 (1976).

\bibitem{Kuzmin} V. Kuzmin, V. Rubakov, M. Shaposhnikov, Phys. Lett. B
{\bf155}, 36 (1985);
P. Arnold, L. McLerran, Phys. Rev. D {\bf37}, 1020 (1988);
N. S. Manton, Phys. Rev. D {\bf28}, 2019 (1983);
F. R. Klinkhammer and N. S. Manton, Phys. Rev. D {\bf30}, 2212 (1984).

\bibitem{Hall1} L.J. Hall and Suzuki, Nucl. Phys. B {\bf231}, 419 (1984);
C. Aulakh and R. Mohapatra, Phys. Lett. B {\bf119}, 136 (1983);
I.H. Lee, Nucl. Phys. B {\bf246}, 120 (1984);
G.G. Ross and J.W.F. Valle, Phys. Lett. B {\bf151}, 375 (1985);
S.Dawson, Nucl. Phys. B {\bf261}, 297 (1985);
F. Zwirner, Phys. Lett. B {\bf132}, 103 (1983);
R. Barbieri, and A. Masiero, Nucl. Phys. B {\bf267}, 679 (1986);
S. Dimopoulos and L. J. Hall, Phys. Lett. B {\bf196}, 135 (1987).


\bibitem{Ibanez} L. E. Ib\'{a}\~{n}ez and Fernando Quevedo, Phys.
Lett. B {\bf283}, 261 (1992).

\bibitem {cline} J. Cline, K Kainulainen, K. A.
Olive, University of Minnesota preprint, UMN-TH-1201/93 (1993);

\bibitem{Ross} H. Dreiner and G.G. Ross, University of
Oxford preprint OUTP-92-08P.

\bibitem{Barr} S.M. Barr, Phys. Rev. D {\bf44}, 3062 (1991);
 D.B. Kaplan, Phys. Rev. Lett. {\bf68}, 741 (1992).

\bibitem{Dod} S. Dodelson, B.R. Greene,
L.M. Widrow, Nuc. Phys. B {\bf372}, 467 (1992).

\bibitem{Sad} D.O. Caldwell {\em et al.}, Phys. Rev. Lett. {\bf61}, 510 (1988).

\bibitem{Ahlen} S.P. Ahlen {\em et al.}, Phys. Lett. B {\bf195}, 603 (1987).

\bibitem{Kolb} E.W. Kolb and M. S. Turner, The Early Universe
(Addison-Wesley, Redwood City, CA, 1990) and references therein.

\bibitem{Lindley} D. Lindley, Astrophys. J. {\bf294}, 1 (1985);
J. Audouze, D. Lindley and  J. Silk, Astrophys. J. {\bf293}, L53 (1985).

\bibitem{Ellis} J. Ellis, G. Gelmini, C. Jarlskog, G.G.
Ross and J.W.F. Valle, Phys.
Lett. B {\bf150}, 142 (1985).

\bibitem{Three}S. M. Barr, R. S. Chivukula and E. Fahri, Phys. Lett.
B {\bf241}, 387 (1991);
S. M. Barr, Phys. Rev. D {\bf44}, 3062 (1992);
D. Kaplan, Phys. Rev. Lett. {\bf68}, 741 (1992).








\end{thebibliography}
\end{document}